# OPTIMUM PARTITION PARAMETER OF DIVIDE-AND-CONQUER ALGORITHM FOR SOLVING CLOSEST-PAIR PROBLEM

Mohammad Zaidul Karim[1] and Nargis Akter[2]

[1]Department of Computer Science and Engineering, Daffodil International University, Bangladesh

`mzkarim@daffodilvarsity.edu.bd`

[2]Department of Computer Science and Engineering, Institute of Science and Technology, National University of Bangladesh

`luckytahseen@hotmail.com`

*ABSTRACT*

*Divide and Conquer is a well known algorithmic procedure for solving many kinds of problem. In this procedure, the problem is partitioned into two parts until the problem is trivially solvable. Finding the distance of the closest pair is an interesting topic in computer science. With divide and conquer algorithm we can solve closest pair problem. Here also the problem is partitioned into two parts until the problem is trivially solvable. But it is theoretically and practically observed that sometimes partitioning the problem space into more than two parts can give better performances. In this paper, a new proposal is given that dividing the problem space into (n) number of parts can give better result while divide and conquer algorithm is used for solving the closest pair of point's problem.*

## 1.INTRODUCTION

The Closest-Pair problem is considered an easy problem, in the sense that there are a number of other geometric problems (e.g. nearest neighbors and minimal spanning trees) that find the closest pair as part of their solution [1]. This problem and its generalizations arise in areas such as statistics, pattern recognition and molecular biology.

At present time, many algorithms are known for solving the Closest-Pair problem in any dimension k > 1, with optimal time complexity [2]. The Closest-Pair is also one of the first non-trivial computational problems that were solved efficiently using the divide-and-conquer strategy and it became since a classical, textbook example for this technique.

During studying, we find out the optimal parameter for divide-and-conquer algorithm. Generally two is used as partition parameter but we theoretically and practically observed that two is not the best parameter of divide-and-conquer algorithm for solving famous closest pair problem. In this circumstance we proposed (n) partition parameter is the better optimum partition parameter for divide-and-conquer algorithm to solve closest pair problem.

## 2. DIVIDE AND CONQUER

In computer science, divide and conquer (D&C) is an important algorithm design paradigm based on multi-branched recursion. A divide and conquer algorithm works by recursively breaking down a problem into two or more sub-problems of the same or related type, until these become simple enough to be solved directly. The solutions to the sub-problems are then combined to give a solution to the original problem. The most well known algorithm design strategies are:

- Divide instance of problem into two or more smaller instances
- Solve smaller instances recursively
- Obtain solution to original instance by combining these solutions

## 3. CLOSEST-PAIR PROBLEM

The closest pair of point's problem or closest pair problem is a problem of computational geometry. The closest pair and its generalizations arise in areas like statistics, pattern recognition and molecular biology. The closest pair problem is defined as follows: Given a set of points, determine the two points that are closest to each other in terms of distance. Furthermore, if there is more than one pair of points with the closest distance, all such pairs should be identified. So the input is a point set with size $n$ and the output is a point pair set.

Its two-dimensional version, for points in the plane, was among the first geometric problems which were treated at the origins of the systematic study of the computational complexity of geometric algorithms.

A naive algorithm of finding distances between all pairs of points and selecting the minimum requires $O(dn^2)$ time. It turns out that the problem may be solved in $O(n \log n)$ time in a Euclidean space of fixed dimension d.

### 3.1 Closest Pair in the Plane in two dimensions

Given a set of points $S$ in the plane, we partition it into two subsets $S_1$ and $S_2$ by a vertical line $l$ such that the points in $S_1$ are to the left of $l$ and those in $S_2$ are to the right of $l$.

We now recursively solve the problem on these two sets obtaining minimum distances of $d_1$ (for $S_1$), and $d_2$ (for $S_2$). We let $d$ be the minimum of these. Now, if the closest pair of the whole set consists of one point from each subset, and then these two points must be within $d$ of $l$. This area is represented as the two strips $P_1$ and $P_2$ on either side of $l$ (as shown in the Figure 1).

International Journal of Computer Science & Information Technology (IJCSIT) Vol 3, No 5, Oct 2011

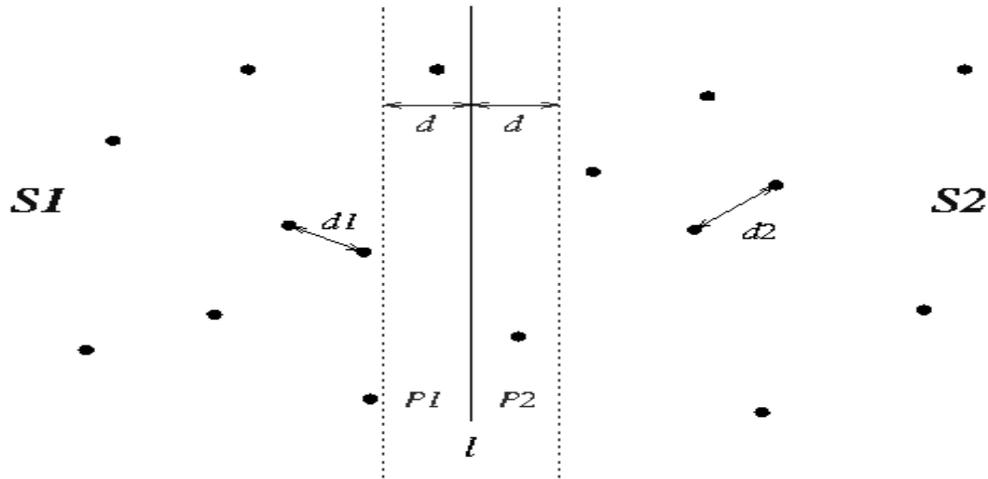

Figure: 1   Divide-and-conquer in two dimensions

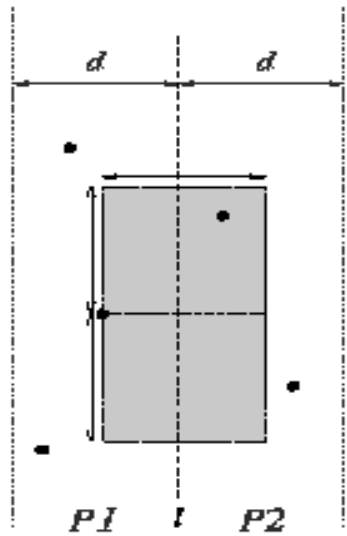

Figure: 2   Strip View

If the closest pair straddles the dividing line, then each point of the pair must be within $d = \min(d_1, d_2)$ of the dividing line.

In this situation for any particular point $p$ in one strip, only points that meet the following constraints in the other strip need to be checked:

- those points within $d$ of $p$ in the direction of the other strip
- those within $d$ of $p$ in the positive and negative $y$ directions

Simply because points outside of this bounding box cannot be less than *d* units from *p* (Figure: 2). It happens because every point in this box is at least *d* apart, there can be at most six points within it. Here we don't need to check all $n^2$ points. We have to do is sort the points in the strip by their *y*-coordinates and scan the points in order, checking each point against a maximum of 6 of its neighbors. This means at most 6*n* comparisons are required to check all candidate pairs. However, since we sorted the points in the strip by their *y*-coordinates the process of merging our two subsets is not linear, but in fact takes O(*n*log*n*) time.

### 3.2 Closest Pair of a set of points:

Divide the set into two equal sized parts by the line *l*, and recursively compute the minimal distance in each part.

a) Let *d* be the minimal of the two minimal distances. It takes O(1) time.
b) Eliminate points that lie farther than *d* apart from *l*. It takes O(*n*) time.
c) Sort the remaining points according to their *y*-coordinates. This Step is a sort that takes O(*n*log*n*) time.
d) Scan the remaining points in the *y* order and compute the distances of each point to its five neighbors. It takes O(*n*) time.

e) If any of these distances is less than *d* then update *d*. It takes O(1) time

Steps define the merging process must be repeated log*n* times because this is a divide and conquer algorithm.

## 4. CLOSEST PAIR ALGORITHM

Let P be a set of n >1 points in the XY-plane. The closest pair in P can be found in O(n log n) time using the divide-and-conquer algorithm.

### 4.1 Closest-Pair (P)

1  Presort points in P along the x-coordinate.
2  Split the ordered set P into two equal-size subsets by the vertical line l defined by the equation x = x $_{median}$
3  Solve the problem recursively in the left and right subsets. This will give the left-side and right-side minimal distances $d_L$ and $d_R$, respectively.
4  Find the minimal distance $d_{LR}$ among the pair of points in which one point lies on the left of the dividing vertical line and the other point lies to the right.
5  The final answer is the minimum between $d_L$, $d_R$, and $d_{LR}$.

Above Pseudocode for the divide-and-conquer Closest-Pair algorithm, first presented by Bentley and Shamos in 1976.[3]

Since we are splitting a set of n points in two sets of n=2 points each, the recurrence relation describing the running time of the closet-pair algorithm is T(n) = 2T(n=2)+ f (n), where f(n) is the running time for finding the distance $d_{LR}$ in step 4.

At first sight it seems that something of the order of $n^2/4$ distance comparisons will be required to compute $d_{LR}$. However, Bentley and Shamos [3] noted that the knowledge of both distances $d_L$ and $d_R$ induces a scarcity condition over the set P.

Let d = min ( $d_L$, $d_R$ ) and consider the vertical slab of width 2d centered at line l. If there is any pair in P closer than d, both points of the pair must lie on opposite sides within the slab. Also, because the minimum separation distance of points on either side of l is d, any square region of the slab, with side 2d, "can contain at most a constant number c of points" [3], depending on the used metric.

As a consequence of this scarcity condition, if the points in P are presorted by y-coordinate, the computation of $d_{LR}$ can be done in linear time. Therefore, we obtain the recurrence relation T(n) = 2T(n=2)+O(n), giving an O(n lg n) asymptotically optimal algorithm.

## 4.2 Closest Pair Analysis

Let T(n) be the time required to solve the problem for n points:

– Divide: $O$ (1)

– Conquer: 2T(n/2)

– Combine: $O$ (n)

The precise form of the recurrence is: $T(n) = T(\lceil n/2 \rceil) + T(\lfloor n/2 \rfloor) + O(n)$
Final recurrence is $T(n) = 2T(n/2) + O(n)$, which solves to $T(n) = O(n \log n)$.

## 5. NEW APPROACH OF DIVIDE-AND-CONQUER ALGORITHM

In the previous sections, the divide and conquer approach to solve the closest pair problem was shown. In that approach the problem space was partitioned into two different problem spaces until the problem is trivially solvable. Here, we propose a different approach where the problem space is divided into (n) different problem spaces instead of the ordinary approach of dividing the problem space into two. After that the supremacy of the new approach over the ordinary approach of divide-and-conquer algorithm on solving the closest pair problem on points is also shown mathematically.

### 5.1 Motivation

With the invention of computers, two-parametric algebra, number system, and graphs among other systems started to flourish with accelerated speed. Boolean algebra got its important applications in computer technology, binary number system has occupied the core of computer arithmetic, and binary trees have become inseparable in mathematical analysis of complexity of algorithms and in the development of efficient algorithms.

Since two has been being used as a parameter having significant influence in the efficiency of the concerned algorithm, the claim of its supremacy over other values should be subject to rigorous verification. Extensive works have already been done on optimality of ternary trees[4] and their VLSI embedding[5]. K-ary partitions, k-ary trees have received attention recently. G¨obel et al. [4] show supremacy of ternary trees. Schaffer and Sedgwick [6] and Kaykobad et al [7]. show improvement of heap sort algorithm using ternary trees whereas Paulik[8] votes for the quaternary. Kaykobad et al. [7] also suggest ternary quick sort to enhance its performance.

Megiddo[9] has placed an objection to the standard translation of problems into languages via the binary coding.

With the above work we are inspired that it can be possible to work in the field of divide-and-conquer algorithm, though it divides with binary partition.

Historical background shows that previously no any work that which partition number is the best for divide-and-conquer algorithm. All of the works focuses on central slab of dividing line to minimize comparisons between points. In this perspective we want to find out the optimal partition parameter for Divide-and-conquer algorithm so that its performance grow faster without hampering its complexity O(n lg n) .

## 5.2 New procedure of divide and conquer rule

The new approach of divide and conquer algorithm of dividing the problem space into (n) different spaces is presented here. The total region of points will be divided into (n) numbers of parts where each region has the same number of points. Then a strip will be formed in each of the (n-1) dividing lines as it is done in the ordinary 2 partition algorithm. This approach will be applied recursively but the fact is, for the number of point's n, partition will be (n) numbers (Figure 3). So that in the first attempt of divide the problem then doesn't divides anymore. And this procedure the program doesn't need recursive call.

Now the results will be merged to have the solution of the total problem.

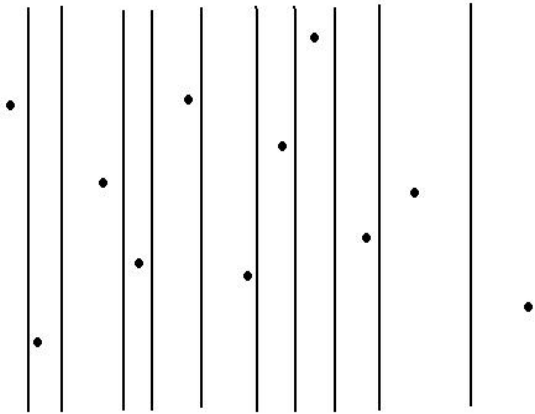

Figure: 3   Number of points n, number of partition (n)

## 5.3 Complexity Analysis

The complexity is analysis is done assuming that there are total n points in the problem region. When the algorithms run on a computer, the most costly process is the distance computation (DC) between points.

$$DC = \sqrt{(x_1 - x_2)^2 + (y_1 - y_2)^2}$$

Where DC is the distance between points $(x_1, y_1)$ and $(x_2, y_2)$ in 2 dimension.
Because for every distance computation, an addition (+), two subtractions (−), two square operations ($^2$) and a square root operation ($\sqrt{}$) is taken place in a 2-dimensional case. Other operations are comparison, which is equivalent to two subtractions (−) operation and data move operation, that are not as costly as the distance computation operation.

### 5.4(n)-Partition Divide and Conquer Approach

In this procedure for n numbers of points local cost is fixed but strip cost varies cause of strip. So our main target is to minimize the local cost as much as we can. If there is n numbers of points and we divide them to (n) numbers of partitions, as a result local cost minimize to zero.

Now in worst-case scenario we count total distance computations cost. Total distance cost is the summation of local distance and strip distance cost. Here only one point remains in every region, so local distance cost is zero.

Again (n-1) numbers of stripes is formed for that n/n point contains in every region.

As we know in the strip calculation most 2 distance computation may be needed for every single point (Jos´e C. Pereira & Fernando G. Lobo 2010) and the points are always divided into (n) regions, there will be at most $\log_n n$ merging steps.

So, total ( $2 * (n(n-1)/n) \log_n n + 0$ ) distance computations will be take place

Where ($2 * (n(n-1)/n) \log_n n$) distance computations are required in the merging step and 0 local distance computations are required. With Corollary 2, (($(a-1)n/a \log_a n + {}^a c_2 (n/a)$)) numbers of distance required for n numbers of points. For partition (2) and (n) local cost is ${}^a c_2 (n/a)=0$. If we put a={ 2,3,4,5,…………,(n-2),(n-1),(n)}; then ( $2(n(n-1)/n) \log_n n$ ) will be the minimum one on worst case scenario where a=n.

### 5.5 Empirical Analysis

Though it is theoretically proved that n partition will be the optimum for n numbers of points on worst case. Now we test it practically. We took 50 random points for this test.

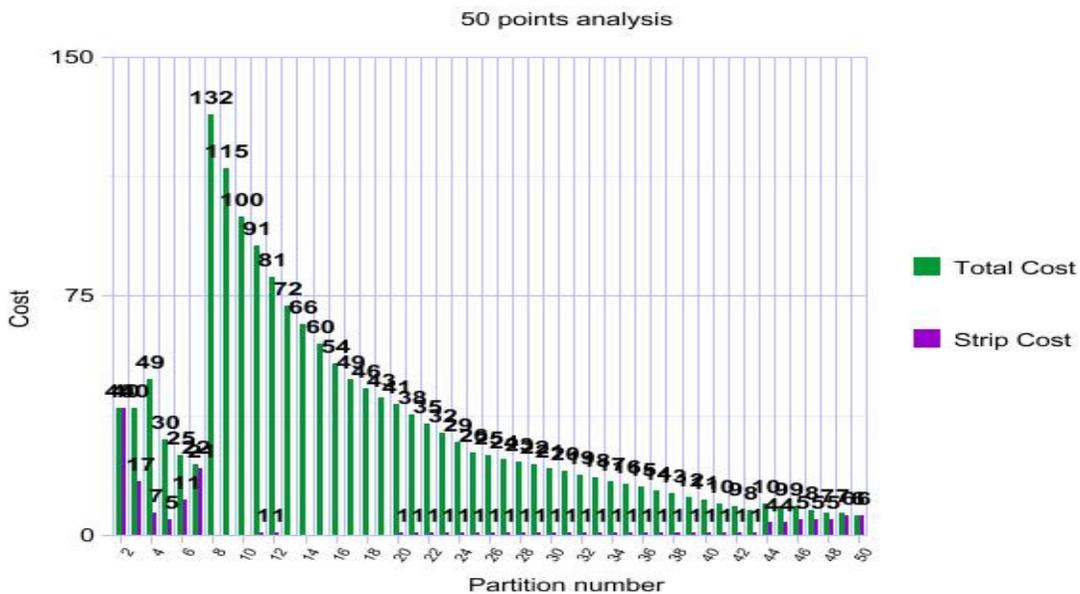

Figure4 : 50 Point Analyses

Figure 4 shows that according to our strategy for the partition number 50 hold the lowest local cost and thus it will be the optimum solution for 50 points.

Now here another question arises that is every time (n) partition went optimum for n numbers of points. For answer we took another test where we randomly generated 50 points 10000times.

Figure: 5 shows that partition 50 appears optimum 5289times out of 10000times. $2^{nd}$ maximum optimum solution is partition 49 and it appears 2160times out of 10000times. $3^{rd}$ maximum optimum solution is partition 48 and it appears 1064times out of 10000times.

So result of experiment 2 supports result of experiment 1. It is clear that for n numbers of points almost more than half of the time partition (n) appears optimum.

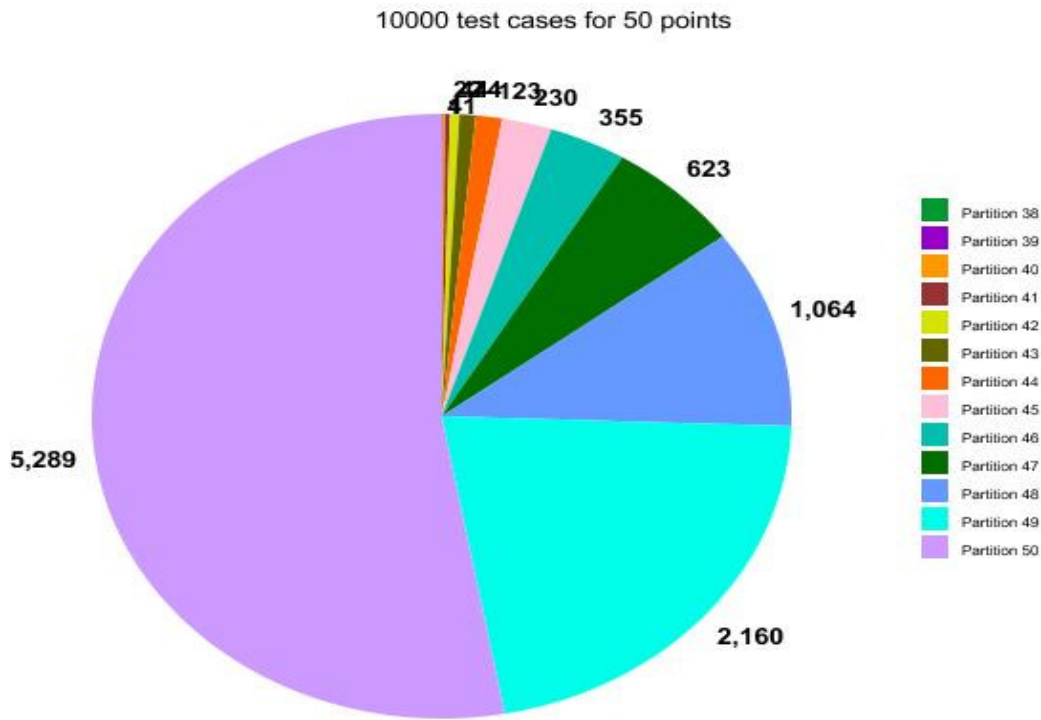

Figure 5 :  10,000 test cases for 50 Points

## 6. CONCLUSION

This paper focus on the use of different parameters instead of the ordinarily used parameter two in algorithms. This paper is organized around two important fields, one is the algorithm Divide and Conquer and the other is the Closest pair of points' problem.

In this paper the comparison is made on the basis of distance computation operation. This is seemed to be the most costly operation takes place in the divide and conquer approach to solve the closest pair of points' problem. Here we try to find which approach uses less distance computations in the worst case, and we find (n)-partition approach uses less distance computations than the other partition approach in the worst case. So, it can be said that (n)-partition approach performs better than other partition approaches.

218

International Journal of Computer Science & Information Technology (IJCSIT) Vol 3, No 5, Oct 2011

## 7. REFERENCES

[1] Bentley, J. L. (1980) , Communications of the ACM 23(4), 214, 1980, p. 226

[2] Smid, M. (2000) , In Handbook of Computational Geometry. (J.-R. Sack and J. Urrutia eds.), pp.877–935, Amsterdam: Elsevier Science

[3] Bentley, J. L. and Shamos, M. I. (1976) , In STOC '76: Proceedings of the 8th annual ACM symposium on Theory of Computing, pp. 220–230, New York, NY, USA: ACM

[4] F. G¨obel and C. Hoede. On an optimality property of ternary trees. Information and Control, Vol. 42, 1979, P.10-26.

[5] S.S. Pinter and Y.Wolfstahl. Embedding ternary treee in VLSI arrays. Znjomxation Processing Letters 26,187-191 (1987).

[6] R. Schaffer and R. Sedgewick. The analysis of heapsort. Journal of Algorithms,Vol. 15, 1993,P. 76 100.

[7] M. Kaykobad, Md. M. Islam, M. E. Amyeen and M. M. Murshed 3 is more promising algorithmic Parameter than 2. Computers Math. Applic. Vol. 36, No. 6,1998, 19-24.

[8] A. Paulik. Worst-case analysis of a generalized heapsort algorithm. Information Processing Letters, Vol. 36, 1990, 159-165.

[9] M. Dietzfelbinger, T. Hagerup, J. Katajainen and M. Penttonen. A reliable ran-domized algorithm for The closest-pair problem. Journal of Algorithms, Vol. 25,1997, 19-51.

[10] Cormen, T. H., Leiserson, C. E., Rivest, R. L., and Stein, C. (2001) , Introduction to Algorithms, MIT Press, 2nd edition

**Authors**

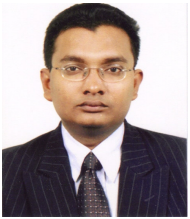

Mohammad Zaidul Karim holds his B.Sc. and M.Sc. in Computer Science from National University, Bangladesh and M.Sc. in Interactive Multimedia Systems from Liverpool John Moores University, UK. Since 2000, he has involved in teaching profession .At present he is working as a senior lecturer in department of Computer Science and Engineering at Daffodil International University. Bangladesh. He is a full professional member of the British Computer Society (MBCS) and Bangladesh Computer Society. He can be reached at: mzkarim@daffodilvarsity.edu.bd

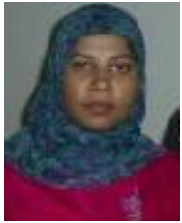

Nargis Akter completed her B.Sc. and M.Sc. in Computer Science from National University, Bangladesh .At present she is working as Assistant Professor at Institute of Science and Technology, National University, Bangladesh. She is holding the position, head of Computer Science and Engineering department. She also holds the position of Joint secretary of the Association of Bangladesh Women's Scientist. Nargis Akter can be reached at: luckytahseen@hotmail.com.